\documentclass[aps,twocolumn,nofootinbib,superscriptaddress,floatfix,prd,letterpaper]{revtex4} 
\usepackage{graphicx,amsmath,amssymb,bm}
\usepackage[usenames,dvipsnames]{xcolor}
\usepackage{xstring}
\usepackage{calligra}
\usepackage{hyperref}

\newcommand{\refeq}[1]{Eq.~(\ref{eq:#1})}          
\newcommand{\refeqs}[2]{Eqs.~(\ref{eq:#1})--(\ref{eq:#2})}

\newcommand{\reffig}[1]{Fig.~\ref{fig:#1}}

\def\mnras{Mon.\ Not.\ R.\ Astron.\ Soc.}

\def\prd{Phys.\ Rev.\ D}

\def\physrep{Phys. Rep.}

\def\jcap{{JCAP\ }}

\newcommand{\iMpch}{\,h\,{\rm Mpc}^{-1}}

\newcommand{\be}{\begin{equation}}
\newcommand{\ee}{\end{equation}}
\newcommand{\bea}{\begin{eqnarray}}
\newcommand{\eea}{\end{eqnarray}}
\newcommand{\vs}{\nonumber\\} 
\def\ba#1\ea{\begin{align}#1\end{align}}

\newcommand{\<}{\langle}
\renewcommand{\>}{\rangle}
\renewcommand{\[}{\left[}
\renewcommand{\]}{\right]}
\renewcommand{\(}{\left(}
\renewcommand{\)}{\right)}

\def\Plin{P_\text{L}}

\newcommand{\avng}{\overline{n}_g}

\renewcommand{\Re}{\operatorname{Re}}
\renewcommand{\Im}{\operatorname{Im}}

\renewcommand{\ln}{\operatorname{ln}}

\newcommand{\cH}{\mathcal{H}}

\renewcommand{\v}[1]{\bm{#1}}

\newcommand{\vk}{\bm{k}}

\newcommand{\vq}{\bm{q}}

\renewcommand{\d}{\delta}

\newcommand{\nhat}{\hat{n}}

\newcommand{\vnhat}{\hat{\v{n}}}

\def\dL{\delta_{\rm L}}
\def\cQ{{\cal Q}}

\def\oursection#1{{\textsc{#1} ---}}

\definecolor{RedWine}{rgb}{0.743,0,0}
\definecolor{RoyalBlue}{rgb}{0.25,.41,.88}

\def\comment#1{}

\begin{document}

\title{The Odd-Parity Galaxy Bispectrum}

\author{Donghui Jeong}
\affiliation{Department of Astronomy and Astrophysics, and Institute
	for Gravitation and the Cosmos,
	The Pennsylvania State University, University Park, PA 16802, USA}

\author{Fabian Schmidt}
\affiliation{Max-Planck-Institut f\"ur Astrophysik, Karl-Schwarzschild-Str. 1, 85748 Garching, Germany}

\begin{abstract}
The galaxy bispectrum contains a wealth of information about the early universe, gravity, as well as astrophysics such as galaxy bias. 
In this paper, we study the parity-odd part of the galaxy bispectrum
which is hitherto unexplored.
In the standard cosmological model, the odd-parity bispectrum is generated by galaxy velocities through redshift-space distortions. While small in the case of General Relativity coupled with smooth dark energy, the signal could be larger in modified gravity scenarios. Thus, apart from being a very 
useful consistency test of measurements of galaxy clustering,
the odd bispectrum offers a novel avenue for searching for new physics.
\end{abstract}

\date{\today}

\maketitle

\oursection{Introduction}
In the standard cosmological model, the $n$-point correlation functions 
of the  matter distribution in the Universe are invariant under the 
parity transformation, or have even parity.
This is because the formation and evolution of large-scale structure is mainly driven by scalar perturbations.
We can, therefore, pursue the signatures of physics beyond the 
standard cosmological model by searching for odd-parity correlation
functions, as suggested by \cite{kamionkowski/souradeep} in the context of 
the cosmic microwave background (CMB) bispectrum.

In this paper, we explore a novel method of measuring the 
parity-violating (odd-parity) part of the correlation functions 
in galaxy clustering. Unlike the case for the CMB where, apart
from instrumental systematics, only  
parity-violating new physics can generate odd-parity correlation 
\cite{kamionkowski/souradeep}, 
the correlation functions of galaxies possess odd-parity parts even in 
the absence of parity-violating new physics.
It is therefore important to quantify their amplitude in order to use the 
odd-parity correlation functions as probes of new physics. Moreover,
we will see that this signal
contains valuable cosmological information even without
parity-violating new physics.

Ref.~\cite{kamionkowski/souradeep} argued that there is no odd-parity bispectrum in three dimensions. However, this only holds in the galaxy rest frame, i.e. if there are no preferred directions; for observed galaxy statistics, the line of sight provides a preferred direction, as the observed redshift is given by $z \simeq \bar z(\chi) + v_\parallel$, where $v_\parallel$ is the galaxy's radial velocity.\footnote{We will adopt the notation of \cite{pbsreview} throughout.}
Thus, there is a non-vanishing \emph{parity-odd galaxy bispectrum}.

\oursection{Imaginary part of galaxy correlation functions}
The reality condition for the galaxy density field is translated into its Fourier
representation as $\d(-\vk)=\d^*(\vk)$. For the three-dimensional power spectrum,
which is defined as
\be
P(\vk_1) \equiv \<\d(\vk_1)\d(\vk_2)\>'
= \<\d(\vk_1)\d(-\vk_1)\>'
= \<|\d(\vk_1)|^2\>'
\,,
\nonumber
\ee
the reality condition implies that the auto power spectrum is a positive-definite real quantity and parity-even: $ P(-\vk) = P(\vk)$.
For the cross power spectrum of two tracers $1$ and $2$, on the other hand,
\be
P_{12}(\vk_1) \equiv \<\d_1(\vk_1)\d_2(\vk_2)\>' = \<\d_1(\vk_1)\d_2(-\vk_1)\>' \,,
\nonumber
\ee
the reality condition only implies $P_{12}(-\vk) = P_{12}^*(\vk)$,
so that one can decompose the cross power spectrum into parity-odd and 
-even parts as
\ba
P_{12}^+(\vk) &= \frac12\[P_{12}(\vk) + P_{12}(-\vk)\] 
= \Re\[P_{12}(\vk)\]
\vs
P_{12}^-(\vk) &= \frac1{2i}\[P_{12}(\vk) - P_{12}(-\vk)\] 
= \Im\[P_{12}(\vk)\]\,.
\ea
This means that the imaginary part of the cross power spectrum probes the 
parity-odd part of the clustering \cite{mcdonald:2009,bonvin/etal:2014}.

Similarly, the reality condition for the bispectrum leads to
$B_{g}(-\vk_1,-\vk_2,-\vk_3)= B^*_{g}(\vk_1,\vk_2,\vk_3)$,
which allows us to decompose the bispectrum into even-parity and odd-parity 
pieces:
\be
B_{g}(\vk_1,\vk_2,\vk_3)
=
B_{g}^+(\vk_1,\vk_2,\vk_3)
+
i B_{g}^-(\vk_1,\vk_2,\vk_3)
\ee
where the parity-even part is real, while the parity-odd part $i B_g^-$ is imaginary.

We see that, among the auto-correlations of any tracer, the bispectrum is the lowest-order statistic that is sensitive to parity.
Including the imaginary, parity-odd part of the bispectrum, which has hitherto been unexplored, in observational analyses means that we \emph{double the number of observables}.
Thus, this is an observable in search of a signal.

\oursection{Cosmological signal in the odd-parity bispectrum}
\begin{figure}
\begin{center}
\includegraphics[width=0.5\textwidth]{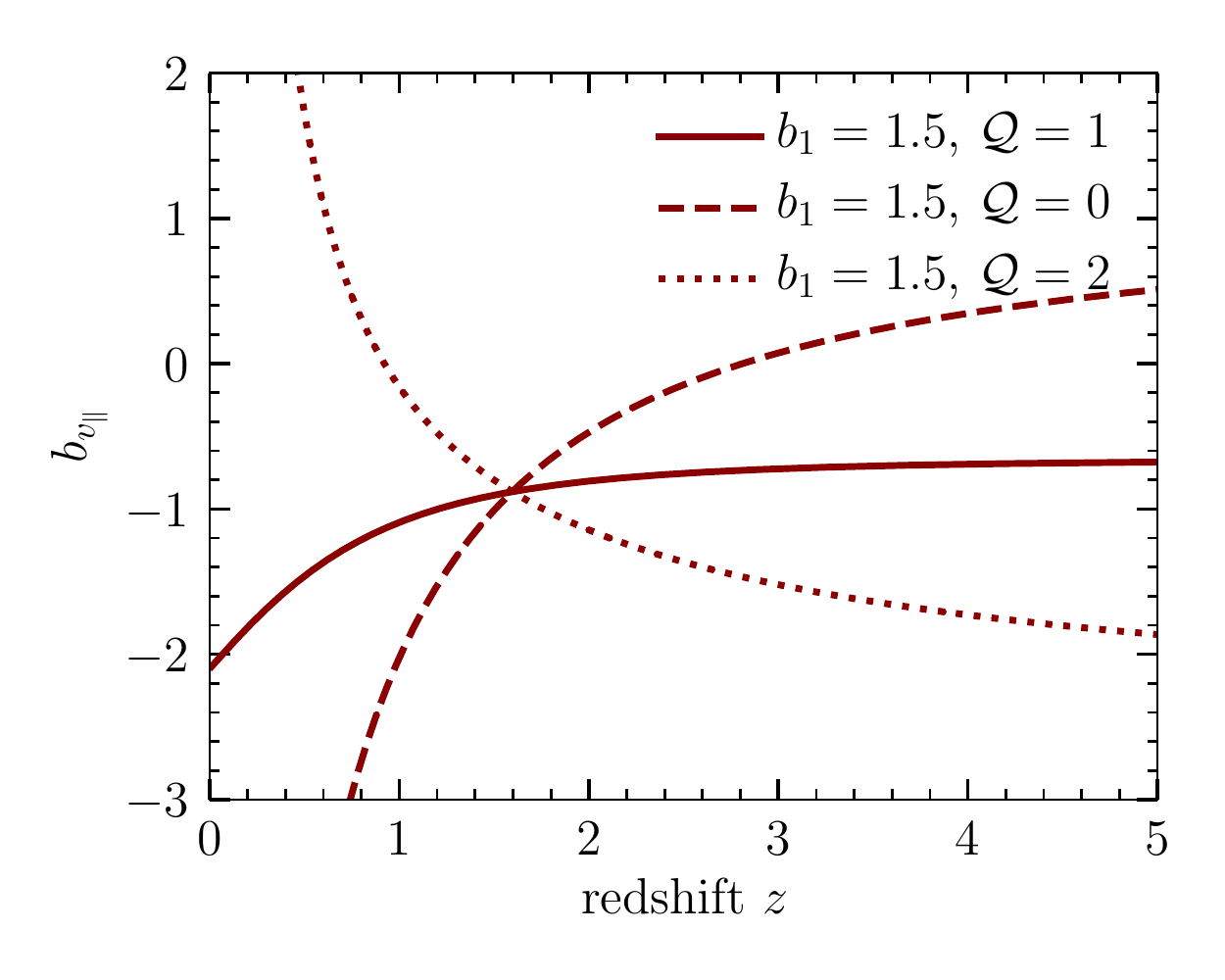}
\caption{%
Evolution of the parameter $b_{v_\parallel}$ for galaxies following 
the universal mass function, $b_e=\delta_c f(b_1-1)$, in the standard 
flat-$\Lambda$CDM cosmology ($\Omega_{\Lambda}=0.69$). 
Except for the $\cQ=1$ case, the $1/(\cH\chi)$ contribution in \refeq{bvp}
diverges at $z=0$.
\label{fig:bvp}%
}
\end{center}
\end{figure}
Due to the peculiar velocity contribution to the observed redshift, the observed galaxy density contrast is distorted from the one in the galaxy rest-frame, 
an effect called \emph{redshift-space distortion} \cite{kaiser:1987}. 
The galaxy density contrast in redshift space (defined with the radial 
distance from the observed redshift $z$) up to second order in perturbations may be written as
\ba
\d_g =\:& 
b_1 \d + b_{v_\parallel} \[v_\parallel (1+b_1\d)\]
+ \frac12 b_2 [\d^2] + b_{K^2} [K^2] \vs%
& 
-\frac{1}{\cH}\partial_\parallel\[v_\parallel\(1+b_1\d\)\]
+
\frac{1}{2\cH^2}\partial_\parallel^2v_\parallel^2\,,
\label{eq:bg}
\ea
where
$b_1,\,b_2,\,b_{K^2}$ are the well-known LIMD (local in matter density) and tidal bias parameters. Note 
that in addition to the usual redshift-space density contrast expression, 
for example in \cite{pkgs_paper}, we include the term proportional to 
$v_\parallel$, the peculiar velocity along the line-of-sight direction.
This term is frequently dropped, as it is suppressed on small scales compared
to terms involving derivatives of the velocity. However, as we will see, 
the term proportional to $v_\parallel$ in \refeq{bg} gives rise 
to the odd-parity bispectrum.

The coefficient $b_{v_\parallel}$ can be derived from the linear-order general relativistic treatment of galaxy clustering 
(e.g., \cite{yoo/etal:2009, challinor/lewis:2011,baldauf/etal:2011,gaugePk,GWpaper,CQGreview}), and is given by
\ba
b_{v_\parallel}
&=
b_e -1 - 2\cQ + \frac{1}{\cH}\frac{dH}{dz} + 2 (\cQ-1)\frac{1}{\cH \chi}
\vs
&=
\frac{d\ln(a^3\avng \chi^2/\cH)}{d\ln a}
-
\cQ
\frac{d\ln (a\chi)^2}{d\ln a}\,.
\label{eq:bvp}
\ea
Here, $\avng$ is the comoving number density
of galaxies and $b_e\equiv d\ln(a^3\avng)/d\ln a$, $\chi$ is the comoving radial
distance to the galaxies, and $\cQ\equiv d\ln \avng/d\ln{\cal M}$ 
parameterizes the change of the observed density contrast due to 
gravitational lensing ($\cal M$ stands for the magnification). 
For a magnitude-limited sample, with the cumulative luminosity function 
$\avng(>L_{\rm min})$, $\cQ=-d\ln\avng(>L_{\rm min})/d\ln L_{\rm min}$ is given
by the slope of the cumulative luminosity function at the limiting luminosity.
All quantities in \refeqs{bg}{bvp} are defined at the observed redshift. 
Note that the two terms in \refeq{bvp} have a straightforward interpretation: the first quantifies the evolution of the mean physical number density of galaxies, while the second does the same for the angular diameter distance squared. Added together, these two contributions in $b_{v_\parallel}$ 
quantify the fractional change of the galaxy number density due to the change in
redshift $\delta z = v_\parallel$ (\reffig{bvp}).

It is worth noting that this simple and physically clear expression for $b_{v_\parallel}$ is only obtained once a proper relativistic calculation is done, which in particular includes terms $\propto \partial_\parallel\Phi$. This latter term is proportional to $v_\parallel$, and hence equally relevant at this order, but has been dropped in quasi-Newtonian calculations of redshift-space distortions such as that in \cite{kaiser:1987} (and many followup papers), leading to a coefficient that differs from, and is not as simple as, \refeq{bvp}.
%
\begin{figure*}[t]
\begin{center}
\includegraphics[width=0.99\textwidth]{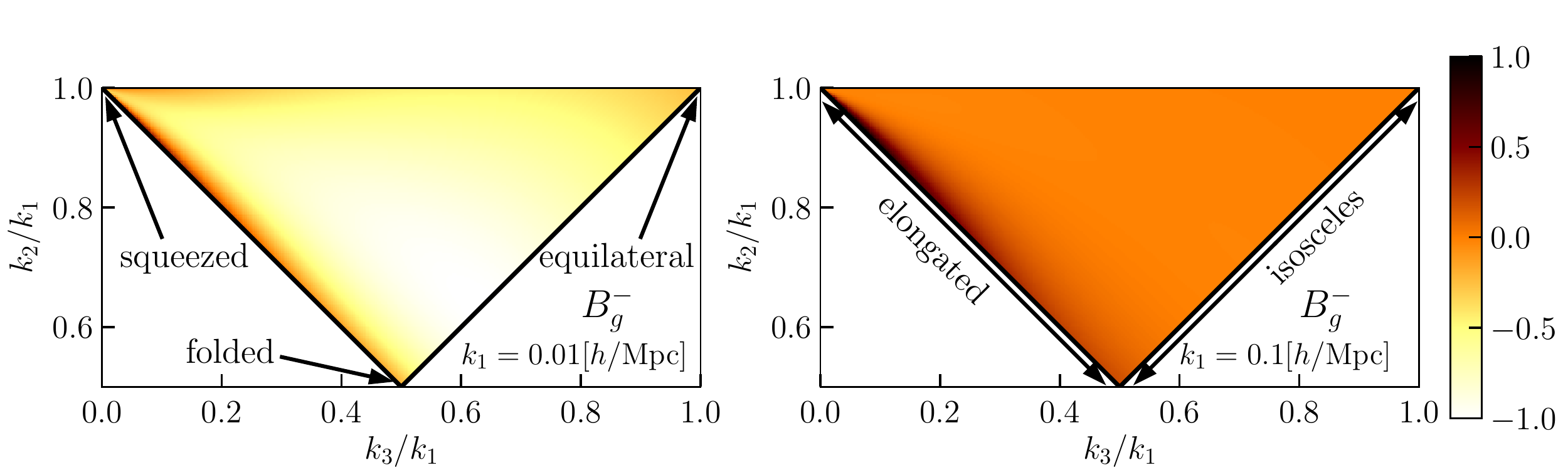}
\caption{%
Shape dependence of the dipole of the odd-parity bispectrum 
$B_g^-(k_1,k_2,k_3)$ for $k_1=0.01~[h/{\rm Mpc}]$ (left) and 
$0.1~[h/{\rm Mpc}]$ (right).
We highlight the shape-dependence by normalizing the amplitude to the
maximum value. In both cases, the odd-parity bispectrum peaks for
elongated ($k_1=k_2+k_3$) configurations. Although the angle average 
complicates the details of the shape-dependence, the basic scale-dependence
is very similar to what is generated in case of the
folded-type primordial non-Gaussianity.
\label{fig:oddBk}%
}
\end{center}
\end{figure*}
    
The leading odd-parity bispectrum comes from the three-point correlations 
involving one power of $b_{v_\parallel} v_\parallel$. It is in fact equivalent
to the bispectrum dipole recently derived in \cite{clarkson/etal:2018}\footnote{Ref.~\cite{clarkson/etal:2018} appeared while this paper was being completed.}
using a full second-order relativistic formalism for galaxy clustering.
Here, we isolate the terms that dominate the signal-to-noise and
correspond to those with the highest number of spatial derivatives.
Correspondingly, in \refeq{bg} we have neglected other relativistic contributions proportional to the gravitational potential, Sachs-Wolfe effect, 
integrated Sachs-Wolfe effect and Shapiro time delay, because their 
contributions are much smaller than those considered here.

In perturbation theory, the line-of-sight velocity field and the density contrast are, to second order, given by
\ba
\frac{v_\parallel(\vk)}{\cH f}
&=
\frac{ik_\parallel}{k^2} 
\biggl[
\dL(\vk)
+
\int_{\vq}
G_2(\vq,\vk-\vq)
\dL(\vq)
\dL(\vk-\vq)
\biggl]\,,
\vs
\delta(\vk)
&=
\dL(\vk)
+
\int_{\vq}
F_2(\vq,\vk-\vq)
\dL(\vq)
\dL(\vk-\vq)\,,
\nonumber
\ea
with the second-order velocity and density kernels
\ba
F_2(\vk_1,\vk_2)
&=
\frac57
+
\frac27\mu_{12}^2
+
\frac12\mu_{12}
\(\frac{k_1}{k_2} + \frac{k_2}{k_1}\)\,,
\vs
G_2(\vk_1,\vk_2)
&=
\frac37
+
\frac47\mu_{12}^2
+
\frac12\mu_{12}
\(\frac{k_1}{k_2} + \frac{k_2}{k_1}\)\,,
\ea
and the linear density contrast field $\dL(\vk)$.
Here, we define $\mu_{ij}$ as the cosine of the angle between two Fourier
vectors $\vk_i$ and $\vk_j$: $\mu_{ij} = \vk_1\cdot\vk_2/(k_1k_2)$.
In Fourier space, the tidal field is related to the density contrast by
$
K_{ij}(\vk) = \(\hat{k}_{i}\hat{k}_j - \frac13 \delta_{ij}\)\d(\vk)\,.
$

The leading-order expression for the odd-parity bispectrum is then,
\begin{widetext}
\ba
\frac{B_g^-(\vk_1,\vk_2,\vk_3)}{b_{v_\parallel} f\cH}
= 2 
&\Biggl\{
\[
b_1 F_2(\vk_1,\vk_2)
+
f \mu_3^2G_2(\vk_1,\vk_2)
+
\frac{b_2}{2}
+
b_{K^2}\(\mu_{12}^2 - \frac13\)
-
b_1
\frac{fk_{3\parallel}}{2}
\(
\frac{\mu_1}{k_1}
+
\frac{\mu_2}{k_2}
\)
+
\frac{f^2k_{3\parallel}^2}{2}
\frac{\mu_1\mu_2}{k_1k_2}
\]
\vs
&\quad\times
\[
\(b_1 + f \mu_1^2\) \frac{\mu_2}{k_2}
+
\(b_1 + f \mu_2^2\) \frac{\mu_1}{k_1}
\]
\label{eq:oddBk}\\
& \  +
\(b_1 + f\mu_1^2\)
    \(b_1 + f\mu_2^2\)
\[
    \frac{\mu_3}{k_3}G_2(\vk_1,\vk_2)
    -
    \frac{b_1}{2}\(\frac{\mu_1}{k_1} + \frac{\mu_2}{k_2}\)
\]
\Biggl\}
\Plin(k_1)\Plin(k_2)
+ 
(2~\mbox{cyclic perm.})\,,
\nonumber
\ea
\end{widetext}
with $\mu_i\equiv \vnhat\cdot\vk_i/k_i$ the cosine of the angle between the 
Fourier-space vector $\vk_i$ and the line-of-sight direction $\vnhat$.
Note that for an order-unity parameter $b_{v_\parallel}$, all contributions to the odd-parity bispectrum are suppressed compared to the leading parity-even 
part of the bispectrum by a factor of $(\cH/k_i)$. 
The scale-dependence of the odd-parity bispectrum is, therefore, very similar
to the case for the ``folded'' primordial bispectrum sourced by initial-state modifications \cite{chen/etal:2007,meerburg/etal:2009,agullo/parker:2010}.
\reffig{oddBk} shows the configuration dependence of the dipole of the 
odd-parity bispectrum at tree level. Like the case for the folded-type
non-Gaussianities, it roughly peaks for degenerate (elongated, or flattened) triangles.

As \refeq{oddBk} is proportional to $\mu_i$, the odd-parity bispectrum 
vanishes when $\vk_i\perp\nhat$ for all three Fourier-space vectors. 
This happens when $\vk_i$ are on the plane perpendicular to the 
line-of-sight direction.
Notice further that the odd-parity bispectrum is free of shot noise, 
and the shot noise only contributes to the covariance matrix.

\oursection{Covariance matrix}
At leading order in perturbation theory, the parity-odd part of the bispectrum is Gaussian distributed with a diagonal covariance matrix given by
\be
\sigma^2[B_g^-(\vk_1,\vk_2,\vk_3)]
=
s_B
\frac{V_{\rm survey}}{2N_t^-}P_g(\vk_1)P_g(\vk_2)P_g(\vk_3),
\label{eq:covBk}
\ee
where $s_B$ is the symmetry factor ($s_B=6$ for equilateral configuration,
$s_B=2$ for isosceles cofigurations, and $s_B=1$ for all others), and 
$N_t^-$ is 
the number of triangles contributing to the estimation of $B_g^-$
within the angular bin $d\mu d\phi$:
\ba
&N_t^-(k_1,k_2,k_3,\mu,\phi) d\mu d\phi
\vs
\simeq&
\,
d\mu d\phi
\,
\left(
    \prod_{i=1}^3 \frac{k_i\Delta k_i}{k_{F_i}^2}
\right)
\times
\left\{
    \begin{array}{cc}
        \pi/2,  & k_i = k_j + k_k\\
        \pi, & {\rm otherwise}
    \end{array}
\right.\,.
\ea
Here, $\Delta k_i$ and $k_{Fi}=(2\pi)/L_{i}$ are, respectively, the width
of the wavenumber bin and the fundamental wavenumber in the $i$-th direction. 
We parameterize the angles between the line-of-sight direction
and the plane of the three wavevectors (the Fourier-space plane 
embedding $\vk_1,\vk_2,\vk_3$) by two parameters: $\mu$, the cosine of the
angle between the line-of-sight direction and the plane, and $\phi$, the 
angle between the $\vk_1$ vector and the line-of-sight direction projected 
onto the plane. 
In this parametrization, the parallel components of all three vectors are 
given as
$k_{1\parallel} = k_1 \mu\cos\phi$, 
$k_{2\parallel} = - k_2 \mu\cos(\alpha+\phi)$, 
$k_{3\parallel} = - k_{1\parallel} - k_{2\parallel}$,
with $\cos \alpha \equiv -\mu_{12}$ being the cosine of the inner angle of the
Fourier-space triangle.

Note that we choose the full angular range 
($\mu\in[-1,1]$ and $\phi\in[0,2\pi]$) and take half of the
available triangular configurations in each angular configuration.
When integrating over the angular configurations ($\mu$ and $\phi$),
$N_t^-(k_1,k_2,k_3)$ is one half of the total number of triangles $N_t(k_1,k_2,k_3)$ (for example, Eq. (4.20) in \cite{pbsreview}).

Using the leading order covariance in \refeq{covBk},
we estimate the signal-to-noise ratio of the odd-parity bispectrum
arising from the redshift-space distortion effect (\refeq{bg})
by using the Fisher information matrix for $b_{v_\parallel}$, which becomes, 
under the null hypothesis, 
\begin{widetext}
\be
\(\frac{S[b_{v_\parallel}]}{N}\)^2
\!\!\equiv 
\(
\frac{b_{v_\parallel}}{\Delta b_{v_\parallel}}
\)^2
\!\!=\!\!
\sum_{(k_1,k_2,k_3)}
\int d\mu \int d\phi
\(\frac{1}{s_BV_{\rm survey}}\)
\frac{\[B_g^-(\vk_1,\vk_2,\vk_3)\]^2}
{P_g(\vk_1) P_g(\vk_2) P_g(\vk_3)}
\left(
    \prod_{i=1}^3 \frac{k_i\Delta k_i}{k_{F_i}^2}
\right)
\times
\left\{
    \begin{array}{cc}
        \pi,  & k_i = k_j + k_k\\
        2\pi, & {\rm otherwise}
    \end{array}
    \right.\,.
    \nonumber
\ee
\end{widetext}
Here, we use the expression for the tree-level galaxy power spectrum in redshift space 
$P_g(\vk_i) = (b_1+f\mu_i^2)^2\Plin(k_i)+ 1/\bar{n}_g$,
including Poisson noise for a survey with mean number density of $\bar{n}_g$.
While we keep all other parameters in \refeq{oddBk} fixed, instead of
marginalizing over them, this is sufficient since they will be constrained
to much greater precision by the parity-even galaxy power spectrum and bispectrum.

We find that 
$\Delta b_{v_\parallel}=2.0$ and $1.4$ (68\% C.L.) for, respectively, the future 
large-scale surveys DESI and Euclid with the survey parameters defined in Table~6 (Sec.~4.1) of \cite{pbsreview}, when including all configurations with $|\vk_i| \leq k_{\rm max}=0.3 \iMpch$.
An ultimate, full-sky cosmic-variance-limited galaxy survey 
in the range $1 \leq z \leq 2$ is expected
to yield $\Delta b_{v_\parallel}\simeq 0.5$. 
With the range of $b_{v_\parallel}$ that we expect from \reffig{bvp}, this
means that the signal-to-noise ratio of detecting the odd-parity bispectrum
for a concordance $\Lambda$CDM cosmology will be 
at most of order unity in any realistic case, 
unless the galaxy number density or luminosity function depend extremely sharply 
on redshift.

This conclusion, however, relies on the fiducial $\Lambda$CDM model. In case
gravity is modified, one expects an increase in velocities due to enhanced gravitational forces. Moreover, the effects of modified gravity can in general be scale-dependent if the additional degree of freedom has a finite mass (as is the case, for example, in $f(R)$ gravity and symmetron scenarios). As a toy model, let us consider the case where velocities are enhanced by 
\be
\v{v}(\vk) = \left(1 + R_v^2k^2 \right) \v{v}_{\Lambda\rm CDM}(\vk).
\ee
Here, $R_v$ corresponds to the length scale (inverse mass, or Compton length) associated with the additional degree of freedom. 
We find that using the odd-parity bispectrum, Euclid and DESI will be able to constrain
$b_{v_\parallel} {R_v}^2 \lesssim  150~h^{-2}{\rm Mpc}^2$, while the above-mentioned
full-sky cosmic-variance-limited survey could achieve $b_{v_\parallel} {R_v}^2 \lesssim 40~h^{-2}{\rm Mpc}^2$. The galaxy power spectrum is in general expected
to yield a tighter constraint on such a scale-dependent modification
of gravity; however, the odd-parity bispectrum is a much cleaner probe, as it
can only be sourced by velocities which are unbiased on large scales
(unlike the galaxy density, and even RSD terms such as $\partial_\parallel v_\parallel$, which can be biased through selection effects). Thus, the odd-parity
bispectrum could be used as a rigorous cross-check of any signs of
new physics found in the galaxy power spectrum.

\oursection{Conclusion}
In this paper, we have studied the imaginary part of the 
galaxy bispectrum that is the lowest-order correlation function sensitive
to the odd-parity component in the galaxy density field. In $\Lambda$CDM cosmological models without parity-violating primordial perturbations, only galaxy velocities
generate the odd-parity bispectrum through redshift-space distortions.

A related complementary approach is to correlate galaxies with an observable that is itself parity-odd like the velocity $v_\parallel$. One such
observable is the kinetic Sunyaev-Zel'dovich (kSZ) effect, which is proportional
to the line-of-sight momentum in ionized gas. The resulting odd-parity galaxy-galaxy-kSZ bispectrum

was recently explored by \cite{smith/etal:2018}.
Here, we have studied the observational prospects for this guaranteed signal
in the odd galaxy bispectrum and
showed that the corresponding signal-to-noise ratio is at most of order unity 
even for idealistic (cosmic-variance limited, full-sky) galaxy surveys,
which is smaller than that reported for the kSZ signal in \cite{smith/etal:2018}. 
This pessimistic outlook changes however if modifications of gravity affect galaxy velocities
significantly. While the detection prospects for such effects are generally
higher in lower-order statistics such as the galaxy power spectrum, the
odd galaxy bispectrum is an exceptionally clean probe since it can only be
sourced by parity-odd terms such as velocities. 
Thus, the odd-parity galaxy bispectrum clearly deserves further attention.

\oursection{Acknowledgments}
DJ acknowledges support from National Science Foundation grant (AST-1517363)
and NASA ATP program   (80NSSC18K1103).
FS acknowledges support from the Starting Grant (ERC-2015-STG 678652) ``GrInflaGal'' of the European Research Council.

\def\eprinttmppp@#1arXiv:@{#1}
\providecommand{\arxivlink[1]}{\href{http://arxiv.org/abs/#1}{arXiv:#1}}
\providecommand{\arxivlinknopre[1]}{\href{http://arxiv.org/abs/#1}{#1}}
\providecommand{\eprintmod}[1][XXXX.XXXX]{\IfSubStr{#1}{arXiv}{\arxivlinknopre{#1}}{\arxivlink{#1}}}
\providecommand{\adsurl}[1]{\href{#1}{ADS}}

\end{document}